\documentclass[apj]{emulateapj}

\usepackage{apjfonts}
\usepackage{multirow, url}
\usepackage{graphicx, amsmath, amssymb, natbib}

\newcommand{\lum}{\,erg\,s$^{-1}$}

\newcommand{\cm}{\,cm$^{-2}$}
\newcommand{\nh}{$N_\mathrm{H}$}

\slugcomment{Submitted to \emph{ApJ} on Jan 10, 2014; Accepted on Mar 11, 2014}

\shorttitle{A Near Chandrasekhar Limit WD in M31}
\shortauthors{Tang et al.}

\begin{document}

\title{An Accreting White Dwarf near the Chandrasekhar Limit in the Andromeda Galaxy}

\author{Sumin Tang\altaffilmark{1,2}, Lars Bildsten\altaffilmark{1,3}, William M. Wolf\altaffilmark{3},  K.~L. Li\altaffilmark{4},  Albert K.~H. Kong\altaffilmark{4},
Yi Cao\altaffilmark{2}, S. Bradley Cenko\altaffilmark{5,6}, Annalisa De Cia\altaffilmark{7}, Mansi M. Kasliwal\altaffilmark{8,13}, Shrinivas R. Kulkarni\altaffilmark{2}, 
Russ R. Laher\altaffilmark{9}, Frank Masci\altaffilmark{10}, Peter E. Nugent\altaffilmark{11, 12}, Daniel A. Perley\altaffilmark{2, 13}, Thomas A. Prince\altaffilmark{2}, and Jason Surace\altaffilmark{9}
}

\altaffiltext{1}{Kavli Institute for Theoretical Physics, University of California, Santa Barbara, CA 93106, USA}
\altaffiltext{2}{Division of Physics, Mathematics, \& Astronomy, California Institute of Technology, Pasadena, CA 91125, USA}
\altaffiltext{3}{Department of Physics, University of California, Santa Barbara, CA 93106, USA}
\altaffiltext{4}{Institute of Astronomy and Department of Physics, National Tsing Hua University, Hsinchu 30013, Taiwan}
\altaffiltext{5}{Astrophysics Science Division, NASA Goddard Space Flight Center, Mail Code 661, Greenbelt, MD 20771, USA}
\altaffiltext{6}{Joint Space Science Institute, University of Maryland, College Park, Maryland 20742, USA}
\altaffiltext{7}{Department of Particle Physics and Astrophysics, Weizmann Institute of Science, Rehovot 76100, Israel}
\altaffiltext{8}{The Observatories, Carnegie Institution for Science, 813 Santa Barbara Street, Pasadena, CA 91101, USA}
\altaffiltext{9}{Spitzer Science Center, California Institute of Technology, M/S 314-6, Pasadena, CA 91125, U.S.A.}
\altaffiltext{10}{Infrared Processing and Analysis Center, California Institute of Technology, Pasadena, CA 91125, USA}
\altaffiltext{11}{Computational Cosmology Center, Lawrence Berkeley National Laboratory, 1 Cyclotron Rd., Berkeley CA 94720, USA}
\altaffiltext{12}{Department of Astronomy, University of California, Berkeley, California 94720-3411, USA}
\altaffiltext{13}{Hubble Fellow}

\begin{abstract}
The iPTF detection of the most recent outburst of the recurrent nova
system RX J0045.4+4154 in the Andromeda Galaxy has enabled the unprecedented study of a
massive ($M>1.3\ M_\odot$) accreting white dwarf (WD). We detected this
nova as part of the near daily iPTF monitoring of M31 to a depth of
$R\approx 21$\,mag and triggered optical photometry, spectroscopy and soft
X-ray monitoring of the outburst. 
Peaking at an absolute magnitude of $M_R=-6.6$ mag, and
with a decay time of 1 mag per day, it is a faint and very fast nova. 
It shows optical emission lines of He/N and expansion
velocities of 1900 to 2600 km s$^{-1}$ 1--4 days after the optical peak.
The {\it Swift} monitoring of
the X-ray evolution revealed a supersoft source (SSS) with $kT_{\rm eff}\approx 90-110
\ {\rm eV}$ that appeared within 5 days after the optical peak, and
lasted only 12 days. Most remarkably, this is
not the first event from this system, rather it is a recurrent nova
with a time between outbursts of approximately 1 year, the shortest
known. Recurrent X-ray emission from this binary was detected by ROSAT
in 1992 and 1993, and the source was well characterized as a $M>1.3\ M_\odot$ WD SSS. 
Based on the observed recurrence time between different outbursts, the duration and effective temperature of the
SS phase,
MESA models of accreting WDs allow us to constrain the accretion
rate to $\dot M>1.7\times10^{-7}\ {M_{\odot}\ {\rm yr}}^{-1}$ and WD mass $>1.30\  M_{\odot}$.
If the WD keeps $30\%$ of the accreted
material, it will take less than a Myr to reach core densities high enough
for carbon ignition (if made of C/O) or electron capture (if made of
O/Ne) to end the binary evolution. 
\end{abstract}


\section{Introduction}

Classical novae are the observable outcome of unstable thermonuclear
burning on accreting white dwarfs (WDs). In most tight binaries with
WDs of typical masses of $M=0.8\ M_\odot$, the recurrence time is tens
of thousands of years. However, if the accretion rate, $\dot M$, is
high and the WD mass is large, the time between flashes can become
short enough so that the recurrence  can be measured. 
Due to the large accretion rate and insignificant ejection mass loss,
recurrent novae (RNe) have been proposed 
to grow toward the Chandrasekhar limit and could be promising progenitors of Type Ia supernovae \citep{1988ApJ...325L..35S, 2010ApJ...719..474D}.
Within our galaxy, these recurrent novae (RNe) have, at the minimum, a recurrence
time of 10 years \citep{2010ApJS..187..275S}. 

As the nearest large galaxy neighbor of the Milky Way,
the Andromeda galaxy (M31) provides the best opportunity for studies of classical and recurrent novae.
Extensive photometric and spectroscopic surveys have been conducted to search for novae in M31,
and have resulted in the discovery of over 900 novae, 
with over 100 having well-sampled light curves, 
optical spectra, or X-ray observations
\citep[see also the website maintained by Pietsch\footnote{\url{http://www.mpe.mpg.de/{\textasciitilde{}}m31novae/opt/m31/M31\_table.html}} and references therein]{1929ApJ....69..103H, 1956AJ.....61...15A, 2004MNRAS.353..571D, 2010A&A...523A..89H, 2011ApJ...734...12S, 2012ApJ...752..133C}.
There are six confirmed RNe in M31, and a few other strong candidates \citep{2013arXiv1307.2296S}.

During our nightly monitoring of M31 in the intermediate Palomar Transient Factory \citep[iPTF; hereafter called simply PTF;][]{2009PASP..121.1395L},
we discovered a transient at the location of a known nova in M31 \citep{2013ATel.5607....1T}, 
and confirmed it to be a RN with a recurrence time of 1yr with four optical outbursts detected in PTF from 2009 to 2013.
During the optical novae, only a fraction of envelope is ejected \citep{1974ApJS...28..247S},
while the remaining envelope is expected to continue hydrogen burning.
As the ejected envelope expands,
the ejecta becomes optically thin, 
and a supersoft source (SSS) powered by hydrogen burning is expected to emerge after the optical nova \citep{2005A&A...439.1061S, 2013ApJ...777..136W}.
Recent observational work has now made it clear that all novae have
 an extended supersoft phase whose
duration and temperature depend solely on the WD Mass \citep{2005A&A...442..879P, 2010ApJ...717..739O,  2013arXiv1312.1241H}. 
We show here that this apparent transient 
behavior is an excellent match to the supersoft phase of a recurrent
 nova, and that the optical novae 
must have been missed in 1992, 1993 and 2001 when X-ray outbursts were seen \citep{1995ApJ...445L.125W, 2004ApJ...609..735W}. 
We describe the PTF discovery and optical follow-up observations in \S 2.
Archival optical and X-ray studies are presented in \S 3 and \S 4, respectively.
{\it Swift} observations and analyses are presented in \S 5.
Theoretical modeling is presented in \S 6.
Our conclusion is in \S 7.

\section{PTF Discovery and Optical Follow-up Observations}

On 2013 Nov 27.08 UT\footnote{All times are in UT}, 
we detected a transient at 
$\alpha = 00^{h} 45^{m} 28.89^{s}$, $\delta = 41^{\circ} 54\arcmin 10\farcs2$
with $R=$18.9 mag \citep{2013ATel.5607....1T}
in the nightly monitoring of M31 in the PTF using the 48-inch telescope at Palomar.
It brightened to $R=$18.3 mag on Nov 28.08. 
No source was detected at the same location to $R<21$ mag in PTF images taken on Nov 26.08 and Nov 25.29. 
There was no detection in 270 nightly PTF $R$-band images taken between 2013 May 19 to Nov 13 to a similar depth during non-bright time.
The transient is coincident within measurement uncertainties with the reported positions of three optical novae or novae candidates, i.e. He/N nova M31N2012-10a \citep{cbat2012-10a, 2012ATel.4503....1S}, 
nova candidates M31N2011-10e \citep{cbat2011-10e, 2011ATel.3725....1B}
and M31N 2008-12a \citep{cbat2008-12a}. 
It is also coincident with the position of a {\it ROSAT} recurrent supersoft transient 
RX J0045.4+4154 \citep{1995ApJ...445L.125W},
which was the first discovery of outbursts from this source.
Therefore, we refer to the transient as RX J0045.4+4154 hereafter.

Following our discovery, we initiated rapid photometric and spectroscopic follow-up observations. 
We obtained $Bg'r'i'$ observations on the Palomar 60-inch telescope \citep[P60;][]{2006PASP..118.1396C} on Nov 29, Nov 30, and Dec 1, 
and $VI$ images on the Low Resolution Imaging Spectrometer \citep[LRIS;][]{1995PASP..107..375O} mounted on the Keck I 10\,m telescope  on Dec 4.
Final reduction of the PTF images was performed using a forced-position Point-Spread Function (PSF) photometry pipeline \citep[][Masci et al. in prep.]{2012PASP..124...62O}.
The P60 and LRIS images were reduced using aperture photometry.
We calibrated the LRIS images using the Local Group Survey catalog \citep[LGS;][]{2006AJ....131.2478M}, 
and obtained a refined position for the transient of 
$\alpha = 00^{h} 45^{m} 28.847^{s}$, $\delta = 41^{\circ} 54\arcmin 10\farcs05$
with uncertainty of $0.1\arcsec$ (dominated by systematic uncertainties; the scattering of positions on the 3 LRIS images are $\approx0.01\arcsec$). 
The optical light curve in 2013 is shown in the top panel in Figure 1.

Optical spectroscopic follow-up of RX J0045.4+4154 was undertaken with 
the Deep Imaging Multi-Object Spectrograph \citep[DEIMOS;][]{2003SPIE.4841.1657F} mounted on the Keck II 10\,m telescope on Nov 29, 2013 (1.3 days post peak),
the Kast double spectrograph \citep{kast} on the Shane 3\,m telescope at Lick Observatory on Dec 1 (3.2 days post peak),
and LRIS mounted on the Keck I 10\,m telescope on Dec 2 (4.1 days post peak).
The spectral resolution is 4.2~\AA, 6~\AA\ (blue side) to 11~\AA\ (red side), and 7~\AA, for the DEIMOS, Kast, and LRIS spectrum, respectively.
Spectra were reduced with standard IRAF routines.

The spectroscopic series of RX J0045.4+4154 is shown in Figure 2.
It showed strong emission lines of Balmer series, 
He\,{\sc i}, He\,{\sc ii}, and N\,{\sc iii}.
The observed optical lines place it into the ``He/N" class of \citet{1992AJ....104..725W}.
The spectra are similar to the HET spectrum taken 0.6 days post peak of the 2012 outburst \citep[M31N2012-10a;][]{2012ATel.4503....1S}.
Such similarity suggests that RN outbursts only depend on system parameters like the WD mass and binary properties. 
Compared with the DEIMOS spectrum,
He\,{\sc i} lines weakened significantly in the LRIS spectrum,
suggesting a decrease in the shell ionization during this period.
The FWHM of H$\rm \alpha$ emission line decreased, from $2600$ km s$^{-1}$ in the DEIMOS spectrum, 
to $1900$  km s$^{-1}$ in the LRIS spectrum, suggesting decreased ejecta velocity.  
Such velocity is at the low end of He/N nova \citep{2011ApJ...734...12S},
and is at the low end of RNe \citep[see e.g.][]{2003ApJ...587L..39K, 2010PASJ...62L..37Y}.
  
The galactic extinction along the line of sight of RX J0045.4+4154 is $A_R=0.134$ and $A_g=0.205$ \citep{2011ApJ...737..103S},
which is the lower limit of extinction of the object.
To estimate the line-of-sight extinction contribution from M31 from the LRIS spectrum,
we subtract the continuum, and measure the flux ratio of $\rm H\alpha/H\beta$ to be 3.89.
Assuming case B recombination (optically thick in Ly$\mathrm \alpha$ Lines, optically thin in all other Hydrogen lines), 
the extinction is $A_R=0.65\pm0.23$ and $A_g=1.0\pm0.35$
\citep[uncertainty comes from the range in expected ratios for Case B of 2.76--3.30;][]{2006agna.book.....O},
which corresponds to \nh $=1.5\pm0.5\times10^{21}$\cm\ \citep{1995A&A...293..889P}.
If the nebula is not optically thin in Balmer lines or if the collisional excitation is non-negligible, 
a higher $\rm H\alpha/H\beta$ ratio is expected relative to the case B recombination,
leading to an overestimated extinction.
Therefore, the above dust extinction and hydrogen column density should be regarded as upper bounds.

\section{Archival Optical Observations}
\subsection{Past Outbursts}
Archival work within PTF revealed other outbursts.
We confirmed outbursts in Oct 2012 \citep[He/N nova M31N2012-10a; ][]{cbat2012-10a, 2012ATel.4503....1S} and Oct 2011 \citep[nova candidate M31N2011-10e;][]{cbat2011-10e, 2011ATel.3725....1B},
and identified another un-reported outburst in Dec 2009. 
The transient is also coincident with the reported positions of nova candidate M31N 2008-12a \citep{cbat2008-12a}. 
We also went back to Arp's survey (1956) of M31 novae in the 1950s and found nothing at this  location in his catalog. 

The observed outbursts are listed in Table \ref{tab:list}.
The recurrence time is 341 days from {\it ROSAT} \citep{1995ApJ...445L.125W}, 
and the recurrence times between the five most recent outbursts (2008--2013) are 342, 689, 363, and 404 days, respectively.
Assuming a recurrence time between 330--410 days,
we expect an outburst in 2010 during Oct 29 to Nov 27.
During this time, 
there are 30 PTF images ($R$ band, typical limiting mag $20.5-21$\,mag),
6 images taken by P60 ($g'$ and $r'$ bands, typical limiting mag $21-22$\,mag),
and 6 images taken by MegaCam on the Canada-France-Hawaii Telescope \citep[CFHT;][]{2003SPIE.4841...72B} 
on Oct 31, 2010 in $u'$, $g'$, and $r'$ with 200s to 600s exposures (limiting mag $\approx24$\,mag).
No outburst was detected.
If there is an outburst with light curve similar to the 2013 and 2011 novae,
it will be detected by PTF during $t-t_0=-1$ to $1$ (where $t_0$ is the optical peak time),
by P60 during $t-t_0=-1$ to $1$, and by CFHT/MegaCam during $t-t_0=-1$ to $7$.
Only two possible windows are left if there was a nova in 2010, i.e. Nov 10--12, or Nov 20--27.

The optical light curves of the 2009, 2011, 2012, and 2013 outbursts are shown in Figure 1, and are similar.
We have daily sampled $g'$, and $R$ band light curves covering the rise, peak, and decay of the 2011 and 2013 outbursts.
The derived $t_2$ (time to decay from the peak by 2\,mag) is $t_{2, g} = $2.0 days in the 2011 outburst,
and  $t_{2, R} = $2.1 days in the 2013 outburst, making it among the fastest novae \citep{2012ApJ...752..133C}.
The declines within 2 mag from the peak are more or less linear, and thus linear regression is used to derive $t_2$.
The peak magnitude in the 2013 outburst is $R=18.34\pm0.08$\,mag at ${\rm MJD}=56624.076$ from PTF data.
The peak magnitudes in the 2011 outburst is $g=18.51\pm0.09$\, mag at ${\rm MJD}=55857.137$ from PTF data,
and $R=18.18\pm0.08$\,mag at ${\rm MJD}=55857.121$ from \citet{2011ATel.3725....1B}.

The simultaneous or nearly simultaneous photometric measurements in the 2011 and 2013 outbursts are listed in Table 1.
Assuming an intrinsic effective temperature of $T\approx8200$\,K as observed in typical novae,
the inferred extinction is $A_R\approx0-0.7$ (\nh $<1.6\times10^{21}$\cm).

Adopting an extinction of $A_R=0.45$ and $A_g = 0.7$ (\nh $=1.0\times10^{21}$\cm),
which is consistent with both the Balmer decrement and available colors of the outbursts,
and given the distance modulus of M31 of 24.4 \citep{2010A&A...509A..70V}, 
the peak absolute magnitude of the 2013 and 2011 outbursts is then $M_R=-6.5$\,mag,
and $M_g=-6.6$\,mag, respectively.
With the fastest decline rate (1 mag d$^{-1}$) yet faint peak magnitude among novae,
RX J0045.4+4154 is an extreme outlier in the canonical 
maximum-magnitude-rate-of-decline (MMRD) relation \citep{1995ApJ...452..704D}.
As discussed by \citet{2011ApJ...735...94K}, 
such faint and fast novae can arise from progenitors containing high
accretion rate and relatively massive white dwarfs (thus lower envelope masses at Hydrogen ignition; see e.g. Wolf et al. 2013), 
as we expect for recurrent novae.

\subsection{Archival HST Observations}
Following the candidate progenitor reported in \cite{2013ATel.5611....1W}, we re-analysed the \textit{Hubble Space Telescope} (HST) archival exposures to investigate photometric properties of RX J0045.4+4154 in quiescence. Besides the ACS/WFC data (Filters: F475W and F814W; Date: 2010 August 7) mentioned in the report, we made use of newer ACS/WFC (Filters: F475W and F814W; Date: 2012 January 10) and WFC3/UVIS\&IR (Filters: F110W, F160W, F275W, and F336W; Date: 2011 January 25 and August 31) observations in the analysis. To search for the progenitor, accurate astrometry with precision down to $0.02\arcsec$--$0.07\arcsec$ between our Keck/LRIS image (see \S2) and the HST images were performed based on 5--18 bright reference stars. Within the Keck/LRIS error circle (astrometric uncertainty dominated) of RX J0045.4+4154, a clear source is present in the optical/UV-band HST images but is undetectable in the IR band. We measured magnitudes of the detected source through the \texttt{Dolphot} PSF photometry package \citep{2000PASP..112.1383D} with HST-dedicated parameters suggested in the manual. All calculated photometric measures are listed in Table \ref{tab:hst}. The mean magnitudes for the optical/UV filters are F275W=23.07, F336W=22.96, F475W=24.24, and F814W=23.91\footnote{See Table \ref{tab:hst} for details. }. For the IR bands, we estimated the upper limits by examining the faintest stars detected by the HST exposures, which gives us F110W $>$ 24.44 (exposure: 800s) and F160W $>$ 25.22 (exposure: 1700s). 
Variability is present in all optical/UV bands with amplitudes up to $0.3$ mag for F275W, $0.6$ mag for F336W, $0.5$ mag for F475W, and $0.2$ mag for F814W,
likely from the high accretion rate disk. 
There is a significant change in color between 2010 and 2012 with $\Delta(B-I)\approx-0.3$ suggesting the system was possibly in a different phase of the nova recurrent cycle
or/and at a different accretion rate. 
The results are consistent with an independent study on the same object by \citet{2014arXiv1401.2905D}.

\section{Archival X-ray Data}
In the M31 second ROSAT PSPC survey catalog \citep{2001A&A...373...63S}, a recurrent supersoft X-ray transient RXJ 0045.4+4154, identified by \cite{1995ApJ...445L.125W}, is spatially coincident with the optical RN (offset: $4\arcsec$; uncertainty $13\arcsec$). RXJ 0045.4+4154 showed two distinct supersoft X-ray outbursts of duration $\approx5$ days in February 1992 and January 1993 (340 days apart) \citep[see][figure 2]{1995ApJ...445L.125W}, in which rapid luminosity changes over the 5-day intervals were clearly seen while a comparably bright X-ray source was completely absent before or after the outbursts. \cite{1995ApJ...445L.125W} interpreted the supersoft X-rays as the consequence of thermonuclear burning of hydrogen on the surface of a WD and estimated the WD atmosphere temperature to be $kT_\mathrm{bb}\approx90$~eV by fitting the spectrum with a blackbody model, inferring a peak luminosity (absorption corrected) of $L_X\approx10^{38}$\lum (0.1--2~keV) and a blackbody radius of $R_\mathrm{bb}\approx 4.7\times10^{8}$~cm ($D_\mathrm{M31} = 780$~kpc assumed). From an O \textsc{viii} ionization edge detection at 0.87~keV, RXJ 0045.4+4154 was believed to be a massive WD with  $M>1.3\ M_\sun$ (i.e., $\log(g)>8.75$). 

\cite{2004ApJ...609..735W} reported a \textit{Chandra}/HRC transient detection in Sep 2001, in which the X-ray transient, named n1-85, 
was mostly quiescent ($L_X\leq2\times10^{36}$\lum), but was detected at one epoch with $L_X\approx 6\times10^{37}$\lum \citep[see][Figure 6]{2004ApJ...609..735W}. 
Their reported coordinates of n1-85 is $9\arcsec$ offset from the optical position of RX J0045.4+4154, 
seemingly making an association with RX J0045.4+4154 unlikely.  
However, we registered the \textit{Chandra}/HRC image to the \textit{Swift}/XRT WCS frame to investigate possible association. 
Using four common X-ray sources detected by both the instruments in the field, 
the astrometric corrected offset of n1-85 and RXJ 0045.4+4154 dropped to about $1\arcsec$, 
which is smaller than the positional uncertainty of the XRT position ($1.6\arcsec$), 
hence, we are confident of the association.

\section{{\it Swift} Observations}

\subsection{{\it Swift} XRT and UVOT light curves}

Given the short X-ray outburst duration observed in ROSAT  \citep{1995ApJ...445L.125W}, 
we launched high cadence ToO X-ray and UV observations with the {\it Swift} observatory \citep{2004ApJ...611.1005G}.
Another {\it Swift} campaign was carried out by M. Henze and collaborators \citep{2013ATel.5627....1T} following our optical discovery \citep{2013ATel.5607....1T}.
RX J0045.4+4154 was observed with the {\it Swift} X-ray telescope 
\citep[XRT; ][]{2005SSRv..120..165B}
and the  Ultraviolet/Optical Telescope \citep[UVOT;][]{2005SSRv..120...95R}  
in a series of 36 observations beginning on 2013 Dec 3 (5 days after the optical peak).
Typically 4--6~ks observations were taken in 1--2~ks snapshots each day from Dec 3 to Dec 16,  and 7--8~ks observations once every 3 days from Dec 17 to Dec 23.
The total exposure time is 97.6~ks. 

A highly variable source is detected in both XRT and UVOT
at the optical position of RX J0045.4+4154 \citep[see also][]{2013ATel.5633....1T}.
The enhanced position given by the online {\it Swift} XRT pipeline software of the UK Swift Science Data Centre at the University of Leicester \citep{2007A&A...469..379E}
is 
$\alpha = 00^{h} 45^{m} 28.80^{s}$, $\delta = 41^{\circ} 54\arcmin 08\farcs8$
with a 90\% error of 1.6\arcsec,
which is consistent with the optical position of the nova.
The {\it Swift} XRT light curve of RX J0045.4+4154 given by the same pipeline is shown in black in the top panel of Figure 3.
The first detection in optical is 1 day before the optical peak,
and we have a 5${\rm \sigma}$ upper limit of $R>21$\,mag 2 days before the optical peak.
Hence the onset of the thermonuclear runaway (TNR) is 1-2 days before the optical peak.
Therefore, from the XRT light curve, we measured the turn-on time (defined as the time it takes for the source to emerge as a SSS  after the onset of TNR) of $<$6--7 days, 
and turn-off time (defined as the time it takes for the source to disappear in X-rays after the onset of TNR) of 18--19 days.
Both timescales are among the shortest ones measured for novae \citep{2013arXiv1312.1241H}.

The UVOT data were reduced with the HEASoft V6.15 package and with the calibration files released in January 2013.
 Aperture photometry was performed using UVOTPRODUCT,
 with a 5\arcsec\ radius for the source,
 and an annulus with inner radius 27\arcsec\ 
and outer radius 35\arcsec\ 
for background,
 as recommended by \citet{2008MNRAS.383..627P}. 
 No bright UV source is located in the background annulus.
The resulting UVOT light curves are shown in the bottom panel of Figure 3.
The magnitudes are in the Vega system. 
It showed significant short-time variations, notably 1 mag variations on hourly timescales in the uvw2 filter \citep[$1928\pm657$~\AA;][]{2008MNRAS.383..627P} during the 5 snapshots in the first {\it Swift} observation.
It also showed a relatively monotonical decline on longer timescales (a few days) after the first {\it Swift} observation. 
The long-term variation amplitude during the {\it Swift} campaign is about 2 mag, 1.4 mag and 1 mag in the uvw2,  uvm2 ($2246\pm498$~\AA) and uvw1  ($2600\pm693$~\AA) filters, respectively.  
No significant variation is detected in the u band ($3465\pm785$~\AA).
The relatively smaller variation amplitudes in UV compared with X-ray
are consistent with the scenario of decreasing effective temperature in the late stage of SSS.

\subsection{\it Effective temperature and column density evolution}

We obtained the \textit{Swift} data 
including the corresponding ancillary files from the \textit{Swift} quick look data archive, extracted the level-2 event files to spectra using \texttt{HEASoft} version 6.14, and performed spectral fittings using \texttt{XSPEC} version 12.8.1 with the an absorbed blackbody model (i.e., \texttt{phabs*bbobyrad}).
Given that the X-ray variability is high over the entire supersoft phase, we split the spectrum into parts according to the observing time
to investigate the luminosity and temperature evolutions of the system.
We divided the observations into three groups (Table \ref{tab:swift}) and fitted each with an absorbed blackbody.
Due to a low photon count, we assumed two possible \nh, i.e. $1.0\times10^{21}$\cm\ and $1.5\times10^{21}$\cm, 
to increase the degrees of freedom of the fits.
The assumed \nh\ are consistent with the extinction measured from the Balmer decrement (\nh $<1.5\times10^{21}$\cm) and optical colors (\nh $=0-1.6\times10^{21}$\cm),
as well as the results from an independent study on the same object by \citet{2014arXiv1401.2904H} who estimated a \nh\ of  $1.1-1.6\times10^{21}$\cm\ from the \textit{Swift} data.

As listed in Table \ref{tab:swift}, the fits show significant variabilities in temperature and luminosity and all best-fit parameters are consistent with the ROSAT outbursts.
The limited number of XRT photons and our lack of knowledge of the  metallicity at the stellar photosphere inhibited our use of the more accurate
stellar atmosphere models of \citet{2010arXiv1011.3628R} that were constructed for hot and massive WDs.

\section{High $\mathrm{\dot{M}}$  and $M$ Nova Models}

The short nova recurrence time, rapid evolution as an X-ray source and high surface temperature during the SSS phase all 
consistently point to a high mass WD. To quantify just how large the WD mass must be to explain the observations, 
we undertook an expansion of the recent work of \cite{2013ApJ...777..136W} 
using the Modules for Experiments in Astrophysics (MESA rev.\ 5596; \citealp{2011ApJS..192....3P, 2013ApJS..208....4P}). 
We simulated WDs with $M=$ 1.30, 1.32, 1.34, and 1.36 $M_\odot$ accreting
material with solar composition, focusing on models that yielded the observed
recurrence time of 1~yr, yielding an
accretion rate range of $1.7\times 10^{-7}<\dot{M}/M_\odot \ {\rm
yr^{-1}}<3.3\times10^{-7}$. These model WDs have core temperatures of
$T_c=3\times 10^7$ K, except for the 1.36 $M_\odot$ model, which had
$T_c=6\times 10^7$ K. The value of $T_c$ does not impact the outcome at these
high $\dot{M}$'s due to the heat buffer created by the even hotter helium layer
($T_{\mathrm{He}}\approx10^8\ \mathrm{K}$) \citep{1995ApJ...445..789P, 1998ApJ...496..376C, 1999ApJ...521L..59P, 2005ApJ...623..398Y, 2013ApJ...777..136W}. The
mass loss prescription during the novae events is a super Eddington wind, as
described in \cite{2013ApJ...762....8D} and \cite{2013ApJ...777..136W}. 

These calculations immediately show that the WD in RX J0045.4+4154 has a
mass of at least $1.3\ M_\odot$, as lower mass WDs could not yield a minimum
recurrence time as short as 1 yr. The $\dot M$'s for the $1.32,\ 1.34, $
and $1.36\ M_\odot$ models that yielded 
$t_{\mathrm{recur}}=1\ \mathrm{yr}$ were, respectively, $3.1\times
10^{-7}$, $2.1\times10^{-7}$, and $1.7\times 10^{-7}\ M_\odot\
\mathrm{yr}^{-1}$. These models retained $\approx 30\%$ of the accreted
material through the outburst, yielding effective accretion rates onto the
helium layer of $\dot{M}_{\mathrm{WD}} \approx 9\times 10^{-8}$, $9\times
10^{-8}$, and $6\times 10^{-8}\ M_\odot\ \mathrm{yr}^{-1}$. Assuming this
matter ultimately stays on the WD and that the effective accretion rate remains
constant, these models would evolve to $M=1.37\ M_\odot$ within $5\times 10^5$
years.

We next studied the location of these high mass WDs in the HR diagram during the SSS phase, all of 
which enter the SSS phase after only about 10--20 days from the onset of the TNR. 
As listed in Table~4, the measured $T_{\mathrm{eff}}$ 
from the two absorbed blackbody models are consistent with each other
within uncertainties, though there is a modest discrepancy in the blackbody luminosities.
As shown in Figure~\ref{fig:mesa}, all theoretical tracks are consistent with the
measured $T_{\mathrm{eff}}$ of RX J0045.4+4154, with either \nh $=1.0\times10^{21}$\cm\ or \nh $=1.5\times 10^{21}$\cm. 
However, the 
WDs have different rates of evolution in the HR diagram depending on their mass, as denoted by the solid 
points that denote the elapsed time between each point. For example, the $1.34\ M_\odot$ WD (orange dotted line and triangles) evolves
through the SSS phase in about 15 days, whereas  the $1.36\ M_\odot$ WD is in a SSS phase for only 8 days. 
Clearly, the $1.30\ M_\odot$ WD evolves
far too slowly. So, this comparison again points to WD masses of $1.30<M/M_\odot<1.36$ in our super Eddington wind model, 
with mass loss due to Roche lobe overflow indicating more massive WDs \citep{2013ApJ...777..136W}. 
The turnoff times could be shorter if 
there was substantial mixing of elements heavier than hydrogen (e.g. He, C, O) into the burning layer during the 
TNR \citep{2005A&A...439.1061S}, pointing to WD masses at the lower end, i.e. $M=1.30\ M_\odot$.

\section{Conclusion}
As a recurrent nova, RX J0045.4+4154 has many remarkable features, with the shortest
recurrence time of 1 year, rapid turn-on and turn-off of the stable burning SSS phase, 
and the highest peak effective temperature during the SSS (100--110 eV).
We showed here that these are all remarkably consistent with theoretical models of
hydrogen thermonuclear flashes on a WD with a mass in the range of $1.30<M/M_\odot<1.36$.
Securely identifying such a massive WD is key to our understanding of the larger problem of deciding how
some massive WDs lead to explosions as Type Ia SNe (if the core is carbon rich) or undergo an
accretion-induced collapse (when the core is composed of O/Ne). At the accretion rate inferred from our theoretical calculations, and assuming (from the duration of the supersoft
phase) that about 30\% of the accreted material stays on the WD, it will take less than a million years for the
core density of the accreting WD to reach values adequate for either an unstable carbon ignition or onset of electron capture (if an O/Ne core).
This assumes that mass loss in the inevitable intervening unstable helium flashes is negligible \citep{2004ApJ...613L.129K}.

This remarkable system will certainly undergo additional outbursts, and we hope that the next one (likely in Nov--Dec 2014) will be studied in even more detail,
especially in the X-rays where spectra can certainly reveal much more about the WD mass and surface composition. More detailed theoretical modeling is
certainly justified, including simulations that resolve the state of the accumulating helium layer.

\acknowledgments
This work was supported by the
National Science Foundation under grants PHY 11-25915, AST 11-09174, and AST 12-05574.
Most of the MESA simulations for this work were made possible by the Triton
Resource. The Triton Resource is a high-performance research
computing system operated by the San Diego Supercomputer
Center at UC San Diego.
This research used resources of the National Energy Research Scientific Computing Center, which is supported by the Office of Science of the U.S. Department of Energy under Contract No. DE-AC02-05CH11231.
AKHK is supported by the National Science Council of the Republic of
China (Taiwan) through grant NSC101-2119-M-008-007-MY3.
MMK acknowledges generous support from the Hubble Fellowship and Carnegie-Princeton Fellowship.
We are grateful to the {\it Swift} Team for the superb timely scheduling of the observations and providing data and analysis tools, and to
Bill Paxton for his development of MESA.

{\it Facilities:}  \facility{PO:1.2m (PTF), 1.5m}, \facility{Keck (DEIMOS, LRIS)}, \facility{{\it Swift} (XRT, UVOT)}.

\bibliographystyle{apj}
\bibliography{09hsdbib}

\clearpage

\onecolumngrid

\begin{figure}
\epsscale{1.0}
\plotone{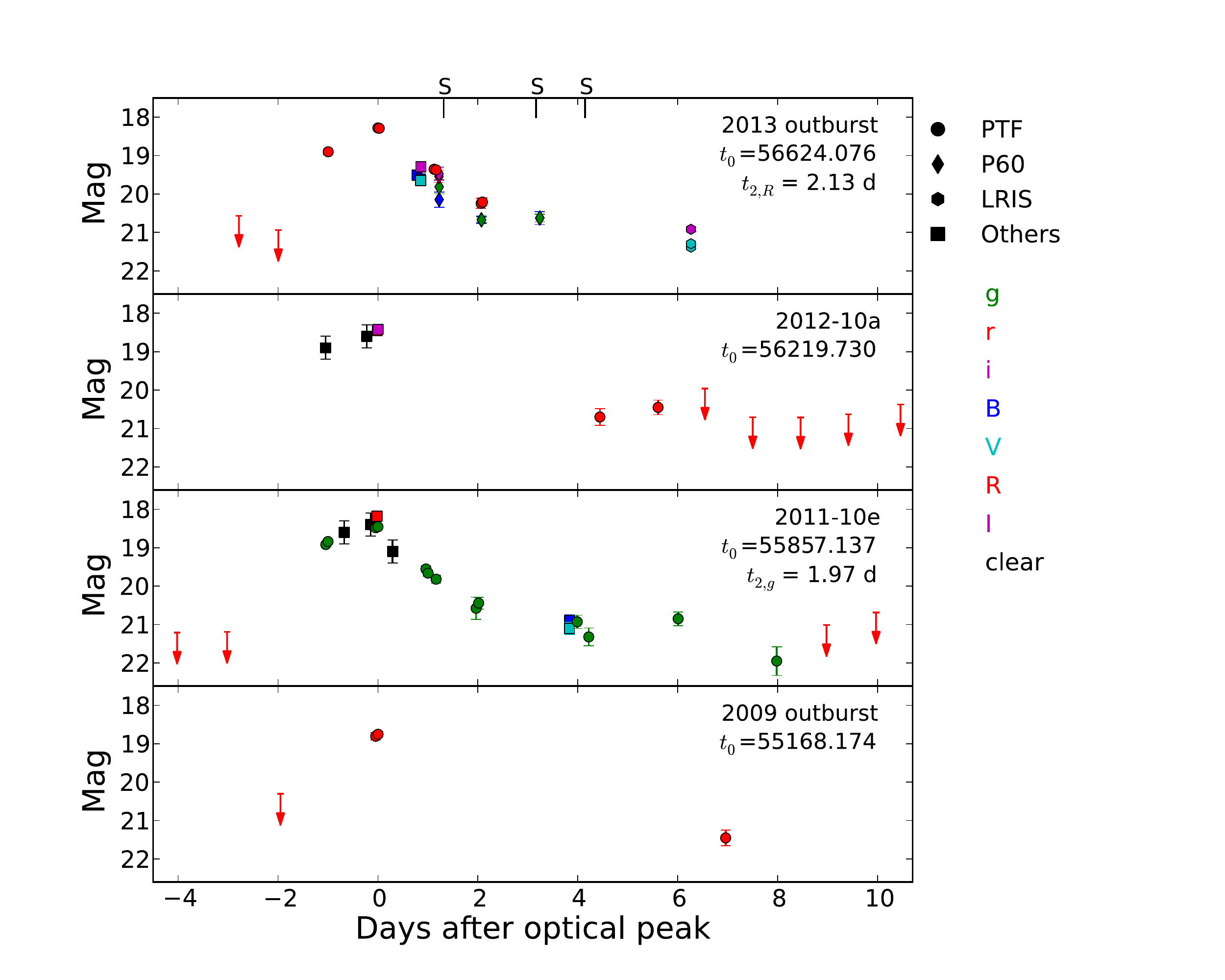}
\caption{Optical light curve of RX J0045.4+4154. Four outbursts were detected in PTF from 2009 to 2013. 
PTF data are shown in circles (green for $g'$ band, and red for $R$ band).
P60 follow-ups are shown in diamonds (blue for $B$ band, green for $g'$ band, red for $r'$ band, and magenta for $i'$ band).
LRIS follow-ups are shown in hexagons (cyan for $V$ band, and magenta for $I$ band).
Published photometry by others \citep{cbat2011-10e, 2011ATel.3725....1B, cbat2012-10a, 2012ATel.4503....1S, 2013ATel.5611....1W} are shown in squares (black for clear, blue for $B$ band, cyan for $V$ band, red for $R$ band, and magenta for $I/i'$ band).
Red arrows are 5$\rm \sigma$ uplimits from PTF R band images. 
On the top axis, the epochs of spectroscopic follow-up of the 2013 outburst are indicated by ``S''.
\label{fig1}}
\end{figure}

\begin{figure}
\epsscale{1.2}
\plotone{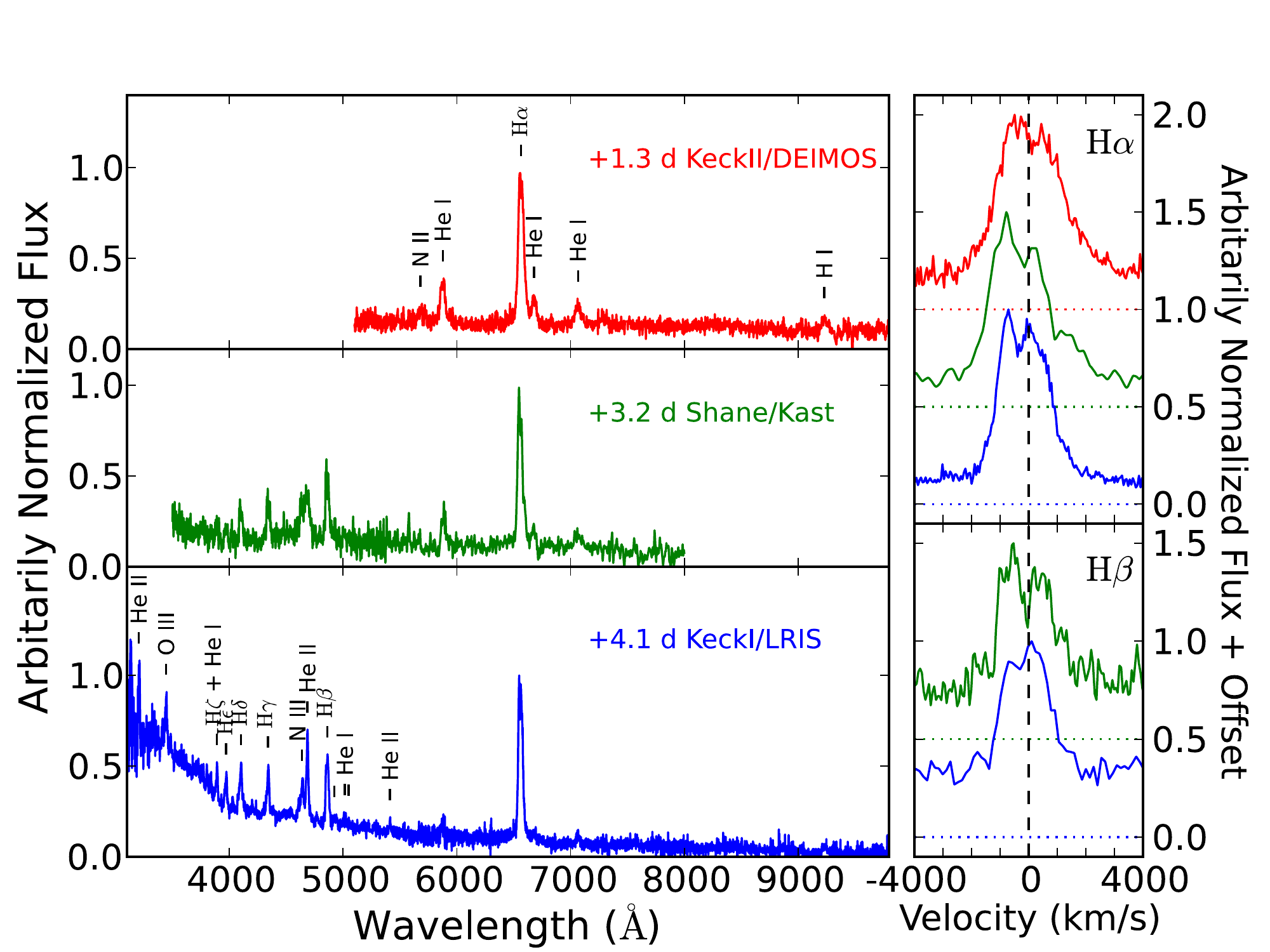}
\caption{Optical spectra of RX J0045.4+4154 during the 2013 outburst. 
{\it Left panels:} arbitrarily normalized spectra. Each spectrum is labelled with the observation date, the telescope and the instrument.
{\it Right panels:} $\rm H\alpha$ and $\rm H\beta$ line profiles.
The Kast spectrum is offseted by +0.5, and the DEIMOS spectrum is offseted by +1.0.
The zero levels of the flux are shown in horizontal dotted lines.
The position of zero velocity is marked by the vertical dashed line.
\label{fig2}}
\end{figure}

\begin{figure}
\epsscale{1.0}
\plotone{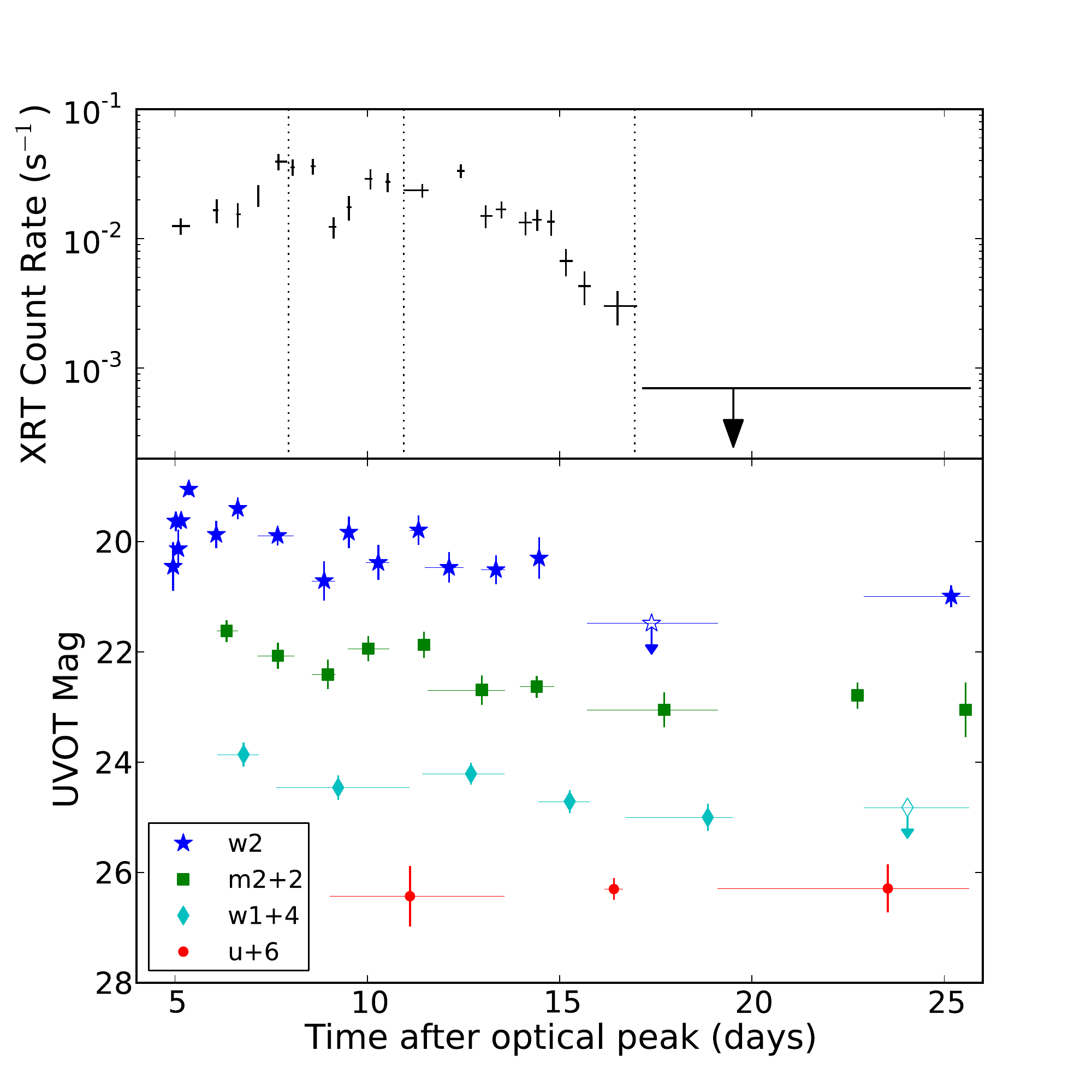}
\caption{{\it Top:} {\it Swift} XRT light curve of RX J0045.4+4154. 
Arrows are 3$\rm \sigma$ upper limits. Vertical dotted lines mark the time bins used in spectral analysis (Figure 4 and Table 4).
{\it Bottom:}  {\it Swift} UVOT light curves of RX J0045.4+4154. The uvm2, uvw1, and u band light curves are shifted by constant values as indicated in the legend.
Open symbols with arrows are 3$\rm \sigma$ upper limits.
\label{fig3}}
\end{figure}

\begin{figure}
\plotone{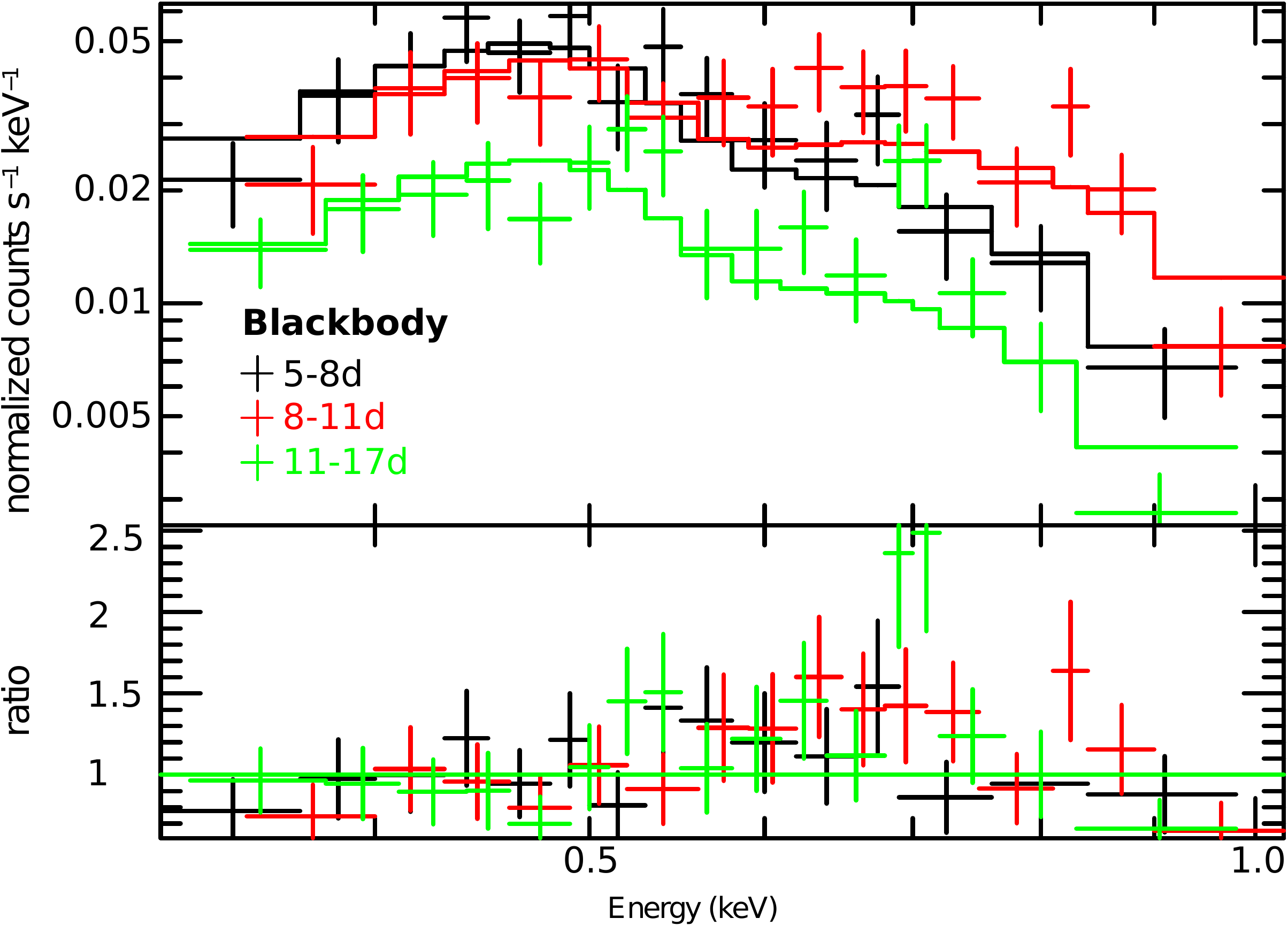}
\caption{The \textit{Swift}/XRT spectra (from 0.3 to around 1~keV) of RX J0045.4+4154 taken at different time bins during the supersoft phase are shown 
with their best-fit black body spectral models assuming a fixed \nh =$1.5\times10^{21}$\cm.
Black for days 5--8 after the optical peak, red for days 8--11, and green for days 11--17.
\label{fig:swift}}
\end{figure}

\begin{figure}
  \centering
  \includegraphics[width = \columnwidth]{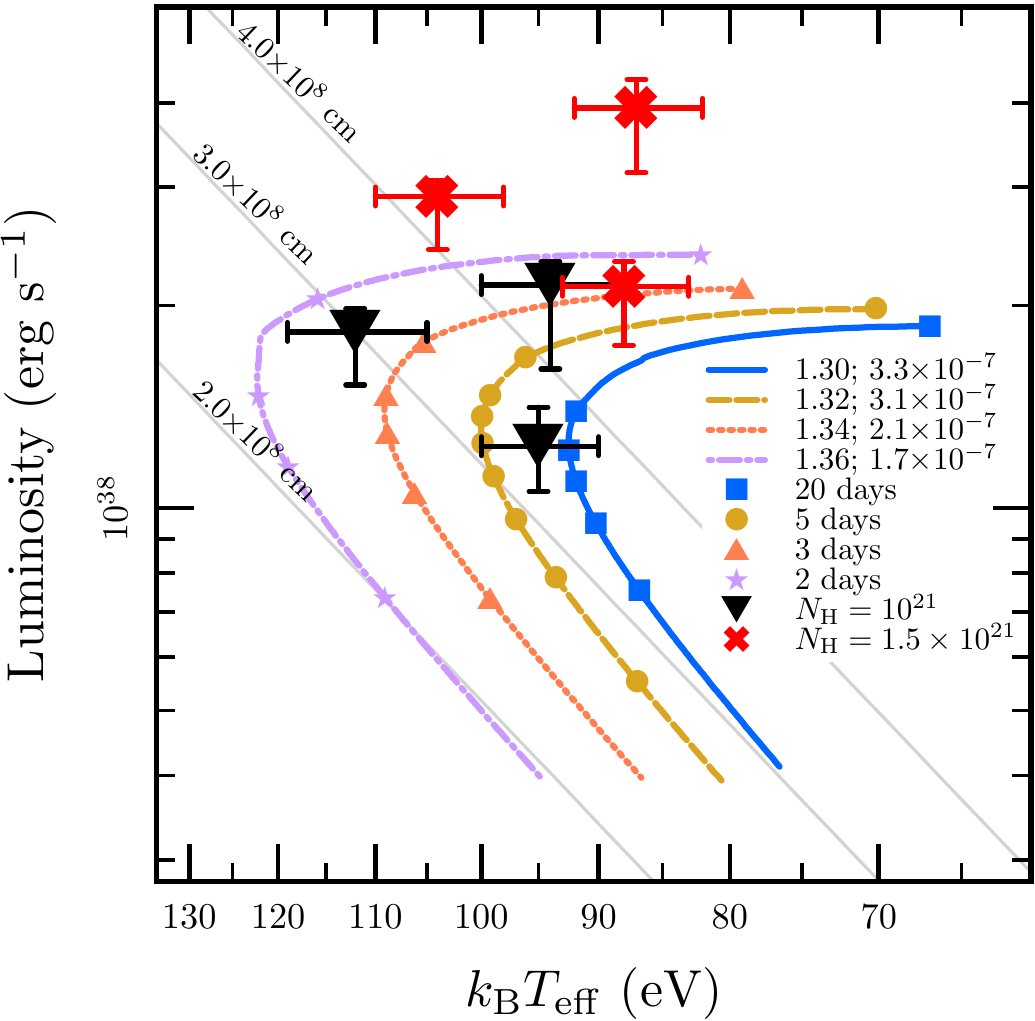}
  \caption{Comparison of observations vs theoretical models on the Luminosity (bolometric) vs $k_BT_{eff}$ plane during the SS phase.
  Best-fit results from the {\it Swift} XRT data using an absorbed blackbody model with fixed \nh =$1.0\times10^{21}$\cm\ 
  and \nh =$1.5\times10^{21}$\cm\  are shown in black upside down triangles and red crosses, respectively,
  with the top-right point from the early stage (5--8 days after optical peak),
  the middle-left point from the middle stage (8--11 days after optical peak),
  and the lower-right point from the late stage (11-17 days after optical peak).
  Note that $L_X(0.3-1\ {\mathrm keV})$ in Table 4 are converted to bolometric luminosity assuming blackbody radiation with best-fit temperatures.
  Evolutionary tracks for the four MESA models
  with $t_{\mathrm{recur}}\approx 1\ \mathrm{yr}$ are shown in thick colored lines with markers;
all the tracks evlove from higher luminosity to lower luminosity. 
  Different lines indicate different WD mass and $\dot{M}$ combinations (indicated in the
  legend in units of $M_\odot$ and $M_\odot\ \mathrm{yr^{-1}}$), and the
  markers separate periods of equal time which are also indicated in the
  legend. More massive stars yield shorter SSS phases due to the smaller
  envelope masses as well as the more vigorous burning
  \citep{2013ApJ...777..136W}. Also shown in gray are lines of constant radius.
}
  \label{fig:mesa}
\end{figure}

\begin{table}[h]
\caption{List of observed outbursts from RX J0045.4+4154.}
\begin{tabular}{ccccl}
\hline
$t_{0, opt}$\footnotemark[1] & $t_{0, X}$\footnotemark[2] & Time since last observed outburst & Source & Reference \\
(UT) & (UT) & (days) &  &  \\
\hline
 & 2013-12-05  & & X-ray ({\it Swift}) & this paper; \citet{2014arXiv1401.2904H} \\
2013-11-28 &   & 404 & Optical (PTF) & \citet{2013ATel.5607....1T}; this paper \\
\hline
2012-10-19 &  & 363 & Optical & \citet{cbat2012-10a, 2012ATel.4503....1S} \\
\hline
2011-10-23 &  & 689 & Optical & \citet{cbat2011-10e, 2011ATel.3725....1B}; this paper \\
\hline
2009-12-03 &  & 342 & Optical (PTF) & this paper \\
\hline
2008-12-26 &  & & Optical &  \citet{cbat2008-12a} \\
\hline
& 2001-09-08  &  & X-ray ({\it Chandra}) & \citet{2004ApJ...609..735W}  \\
\hline
 & 1993-01-11 & 341 & X-ray ({\it ROSAT}) & \citet{1995ApJ...445L.125W} \\
\hline
& 1992-02-05 &  & X-ray ({\it ROSAT}) & \citet{1995ApJ...445L.125W}  \\
\hline
\end{tabular}
\footnotetext[1]{Time of the optical peak.  }
\footnotetext[2]{Time of the X-ray peak. }
\label{tab:list}
\end{table}

\begin{table}[h]
\caption{Optical photometry of RX J0045.4+4154 during the 2013 and 2011 outbursts with (nearly) simultaneous multi-band observations.}
\footnotesize
\begin{tabular}{llcccccccl}
\hline
MJD & $t-t_0$ & B & V & R & I & g' & r' & i' & Ref. \footnotemark[1] \\
& (days) & (mag) & (mag) & (mag) & (mag) & (mag) & (mag) & (mag) & \\ 
\hline
\multicolumn{3}{l}{2013, $t_0=56624.08$ in MJD}\\
\hline
56624.93 & 0.85 & $19.61\pm0.01$ & $19.65\pm0.02$  &&&&& $19.29\pm0.02$& W13\\
56625.26 & 1.18 & $20.15\pm0.20$ & & & & $19.82\pm0.18$ & $19.54\pm0.16$ & $19.47\pm0.17$  & P60 \\
56626.15 & 2.07 & $20.67\pm0.10$ & & $20.32\pm0.16$ & & $20.67\pm0.07$ & & & PTF and P60 \\
56627.32 & 3.24 & $20.63\pm0.17$ & & & & $20.62\pm0.10$ & &  & P60 \\
56630.34 & 6.26 & & $21.30\pm0.02$ & & $20.92\pm0.05$ & & & &  LRIS \\
 \hline
\multicolumn{3}{l}{2011, $t_0=55857.1$ in MJD} \\
\hline
55857.1 & 0 & & & $18.18\pm0.08$ &&$18.51\pm0.09$&&& B11; PTF\\
\hline
\end{tabular}
\footnotetext[1]{References: W13 is Williams et al. 2013; B11 is Barsukova et al. 2011. 
Others are either PTF photometry or our follow-up using Palomar 60-inch or KeckI/LRIS.}
\label{tab:optcolor}
\end{table}

\begin{table}[h]
\centering
\caption{HST Observations of RX J0045.4+4154}
\small
\begin{tabular}{ccrrcc}
\hline
Filter ID & Observing Time  &  Days after\footnotemark[1]  & Days before\footnotemark[1]  & Exposure & Vega Magnitude\\
& (yyyy-mm-dd hh:mm:ss)  & previous nova & next nova & (s) & (mag)\\ 
\hline
F275W & 2011-01-25 04:56:24 & 418 (74) & 271 & 350 & $23.244\pm0.118$ \\
 & 2011-01-25 05:20:52 & 418 (74) & 271 & 660 & $23.078\pm0.072$ \\
 & 2011-08-31 12:18:37 & 636 (292) & 53 & 350 & $23.075\pm0.103$ \\
 & 2011-08-31 12:41:25 & 636 (292) & 53 & 575 & $22.955\pm0.079$ \\
 \hline
F336W & 2011-01-25 04:44:43 & 418 (74) & 271 & 550 & $23.309\pm0.054$ \\
 & 2011-01-25 05:04:54 & 418 (74) & 271 & 800 & $23.118\pm0.040$ \\
 & 2011-08-31 12:06:56 & 636 (292) & 53 & 550 & $22.715\pm0.042$ \\
 & 2011-08-31 12:27:07 & 636 (292) & 53 & 700 & $22.738\pm0.037$ \\
 \hline
F475W & 2010-08-07 12:27:39 & 247 & 431 (87) & 600 & $24.041\pm0.022$ \\
 & 2010-08-07 12:40:20 & 247 & 431 (87) & 370 & $24.041\pm0.029$ \\
 & 2010-08-07 12:49:08 & 247 & 431 (87) & 370 & $24.035\pm0.029$ \\
 & 2010-08-07 12:57:56 & 247 & 431 (87) & 370 & $23.989\pm0.028$ \\
 & 2012-01-10 02:45:00 & 79 & 284 & 700 & $24.448\pm0.028$ \\
 & 2012-01-10 02:59:21 & 79 & 284 & 360 & $24.451\pm0.038$ \\
 & 2012-01-10 03:07:59 & 79 & 284 & 360 & $24.410\pm0.037$ \\
 & 2012-01-10 03:16:37 & 79 & 284 & 470 & $24.478\pm0.033$ \\
 \hline
F814W & 2010-08-07 10:44:08 & 247 & 431 & 350 & $23.846\pm0.045$ \\
 & 2010-08-07 10:52:38 & 247 & 431 (87) & 700 & $23.871\pm0.031$ \\
 & 2010-08-07 11:06:56 & 247 & 431 (87) & 455 & $23.815\pm0.037$ \\
 & 2012-01-10 00:23:51 & 79 & 284 & 350 & $23.981\pm0.049$ \\
 & 2012-01-10 01:09:03 & 79 & 284 & 800 & $23.967\pm0.039$ \\
 & 2012-01-10 01:25:01 & 79 & 284 & 550 & $23.964\pm0.038$ \\
\hline
\\
\end{tabular}
\footnotetext[1]{Days in parenthesis are the ones assuming there was a missing nova on Nov 12, 2010, which is the middle point between the 2009 and the 2011 novae. }
\label{tab:hst}
\end{table}

\begin{table*}[ht]
\centering
\caption{X-ray spectral fitting of RX J0045.4+4154}
\small
\begin{tabular}{cccccc}
\hline
Time\footnotemark[1] & \nh\ & Temperature ($T_\mathrm{bb}$)\footnotemark[2] & Radius\footnotemark[3] ($R_\mathrm{bb}$) & Blackbody emission\footnotemark[4] ($L_X$) & $\chi^2_\nu$ (dof) \\
(days) & ($10^{21}$\cm)  & (eV) & (10$^8$cm) & (10$^{38}$\lum) &  (10$^{38}$\lum) \\ 
\hline
\multicolumn{6}{c}{Single Absorbed Blackbody with a fixed \nh =$1.0\times10^{21}$\cm} \\
5--8 & \multirow{3}{*}{1.0 (fixed)} & $94^{+6}_{-6}$ & $4.7^{+3.3}_{-2.7}$ & $1.2^{+0.1}_{-0.3}$  & \multirow{3}{*}{1.40 (42)}\\
8--11 & & $112^{+7}_{-7}$ & $2.9^{+1.9}_{-1.6}$ & $1.2^{+0.1}_{-0.2}$   \\
11--17 & & $95^{+5}_{-5}$ & $3.4^{+2.3}_{-1.9}$ & $0.7^{+0.1}_{-0.1}$   \\
\hline

\multicolumn{6}{c}{Single Absorbed Blackbody with a fixed \nh =$1.5\times10^{21}$\cm} \\
5--8 & \multirow{3}{*}{1.5 (fixed)} & $87^{+5}_{-5}$ & $7.3^{+5.4}_{-4.3}$ & $2.0^{+0.2}_{-0.4}$  & \multirow{3}{*}{1.27 (42)}  \\
8--11 & & $104^{+6}_{-6}$ & $4.3^{+2.9}_{-2.4}$ & $1.8^{+0.1}_{-0.3}$   & \\
11--17 & & $88^{+5}_{-5}$ & $5.1^{+3.5}_{-2.9}$ & $1.1^{+0.1}_{-0.2}$  & \\ 
\hline



\end{tabular}

\footnotetext[1]{Time after the optical peak in 2013. }
\footnotetext[2]{All uncertainty ranges are 90\% confidence interval. }
\footnotetext[3]{The radii of the blackbody were calculated using the best-fit normalizations (i.e., $\propto R_\mathrm{bb}^2/D_\mathrm{M31}^2$ with $D_\mathrm{M31}=780$~kpc). }
\footnotetext[4]{The luminosities (0.3--1~keV) are absorption corrected with an assumption of $D_\mathrm{M31}=780$~kpc while the uncertainties were estimated by approximating $\Delta\,L_\mathrm{unabs}$/$L_\mathrm{unabs}\approx\Delta\,L_\mathrm{obs}$/$L_\mathrm{obs}$. }
\label{tab:swift}
\end{table*}

\end{document}